%% file: main_FK2.tex
\documentclass[a4paper, 12pt]{article}
\input{preambule.tex}

\author{Boulat Nougmanov}
\title{Optimal squeezing to minimize vulnerability to losses}
\begin{document}
\maketitle
\begin{abstract}
  Non-Gaussian states, described by Wigner quasi-probability distribution taking negative values, are of great interest for various applications of quantum physics. It is known however that they are highly vulnerable to dissipation. In this paper, we show that the robustness of the non-Gaussian states to losses can be significantly improved by pre-squeezing of the quantum state, and find the optimal parameters of the squeezing. As specific examples, we consider such well-known quantum states as Schrodinger cat, Fock , and ``banana'' ones.
\end{abstract}

\section{Introduction}\label{sec:real_intro}

The Wigner function is a quasi-probability distribution {allowing to describe quantum states} in phase space \cite{schleich2015quantum}. "Classical" probability distributions should be non-negative. However, in quantum mechanics, particles can behave as if the probability densities of being in some areas were negative. This does not lead to really negative probabilities in the experiment, since the coordinate and momentum cannot be measured simultaneously \cite{bohr1928quantum}. However, this behavior cannot be interpreted purely classically.

According to Hudson's theorem \cite{hudson1974wigner}, all pure states with a non-negative Wigner function have a Gaussian distribution function, so the terms "nonclassical state" and "non-Gaussian state" are synonymous in this context. { Gaussian states allow a local hidden-variable description in terms of position and momentum,} so all experiments based on Gaussian states and linear (position/momentum) measurements can be efficiently modeled on a classical computer \cite{bell2004speakable, mari2012positive}. Thus, the negativity of the Wigner function is { the  necessary condition of non-classicality of the state. Volume of the negative-valued part of the Wigner function can be used as the quantitative measure of the non-classicality.}

Non-Gaussianity is an important aspect of quantum optics, as it is directly related to the manifestation of nonclassical properties of quantum states of light \cite{walschaers2021non}. This property makes it possible to use non-Gaussian states to implement tasks that are not available in classical optics, such as improving the accuracy of quantum measurements \cite{Gorshenin_LPL_21_065201_2024, Gorshenin_JOSAB_42_425_2025} and implementation of efficient quantum computing protocols.

{ At the same time, it is known that non-Gaussian states are vulnerable to dissipation \cite{Zurek_RMP_75_715_2003}.} In many quantum experiments, losses are relatively small, but as we will see in this paper, the negativity of the state can be lost quite significantly even in the presence of small losses. In this regard, a natural question arises: what can be done to preserve as much negativity as possible?


{ In this work, we propose to use pre-squeezing of the non-Gaussian state for this purpose}. Squeezing of light using $\chi^{(2)}$ crystals is a fairly well-researched procedure in recent decades, which has already found many applications, including interferometry \cite{grangier1987squeezed, xiao1987precision}, gravitational-wave detectors \cite{ligo2011gravitational, aasi2015advanced}, quantum communications \cite{ralph1999continuous} and fundamental tests of quantum mechanics \cite{martin2016bell}. Squeezing does not change the amount of negativity of the Wigner function, but, as we show in this article, the robustness of the state to losses can be significantly improved.


{The article is organized as follows. In Sec.\,\ref{sec:metrics}, we develop a quantitative metrics of vulnerability of a quantum state to losses. In Sec.\,\ref{sec:optim_sq} we find the general form of the squeezing parameters which minimize this metric. In Sec.\,\ref{sec:examples} we apply the developed theory to some specific quantum states used in quantum optics. Finally, in Sec.\,\ref{sec:conclusion}, we summarize the obtained results.}

\section{Metrics of the vulnerability}\label{sec:metrics}

\begin{figure}
\centering
\includegraphics[width=0.25\linewidth]{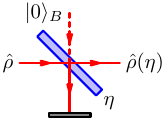}
\caption{Beam splitter as a model for linear losses.}
\label{fig:losses}
\end{figure}

We start by defining losses through an effective beamsplitter model, see Fig.\,\ref{fig:losses}. Following, for example, the work \cite{yuen1978optical}, we introduce a linear interaction of { a mode described by the density operator $\hat{\rho}$ with a heath-bath one in a vacuum state $\ket{0_B}$:
\begin{equation}\label{rho_eta}
  \hat{\rho}(\eta) = \Tr_B\bigl(
    \hat{\mathcal{U}}\,\hat{\rho}\otimes\ket{0_B}\bra{0_B}\,\hat{\mathcal{U}}^\dagger
  \bigr) ,
\end{equation}
Here $\Tr_B$ means {tracing out} the heat-bath subspace and $\mathcal{U}$ is the evolution operators defined by
\begin{equation}
  \hat{\mathcal{U}}^\dagger\hat{a}\hat{\mathcal{U}}
  = \sqrt{\eta}\,\hat{a} + \sqrt{1-\eta}\,\hat{a}_B \,,
\end{equation}
where $\hat{a}$, $\hat{a}_B$ are, respectively, the annihilation operators of the main and the heat-bath modes, and $\eta$ is the quantum efficiency.
}

The Wigner function $W(x, y, \eta)$, corresponding to resulting density matrix $\hat{\rho}(\eta)$, can be obtained as a result of Gaussian blur with the initial Wigner function $W(x, y)$, see Ref.\,\cite{leonhardt1994high}:
\begin{eqw}\label{eq:W_eta}
W(x, y, \eta) =  \int\limits_{-\infty}^{+\infty}\int\limits_{-\infty}^{+\infty} W(\tilde x, \tilde y) \exp\biggl[-\frac{\br{x-\sqrt{\eta}\tilde x}^2+\br{y-\sqrt{\eta}\tilde y}^2}{(1-\eta)}\biggr] \;\frac{d\tilde x \; d\tilde y}{\pi(1-\eta)}
\end{eqw}
As a measure of the degree of non-classicality of light, we use the volume of negativity of the Wigner function \cite{kenfack2004negativity}:
\begin{eqw}\label{eq:VnegDef}
    V_{neg}(W) = -\int\limits_{W(x, p) < 0} W(x, p) dx \; dp
\end{eqw}

{We assume that the losses are relatively small and decompose $V_{neg}$ into a Taylor series in $1-\eta$:}
\begin{eqw}\label{eq:Vneg_by_eta}
    V_{neg}(\eta) = V_{neg}(1)
    - (1-\eta)\frac{\partial V_{neg}}{\partial \eta}\biggr|_{\eta=1}  + \dots
\end{eqw}
It is easy to see that the main metrics for determining the vulnerability of a state is the first derivative of the volume of negativity. In Appendix \ref{Appendix_A} we derive the formula that shows that the sharper the Wigner function in areas of negativity, the faster their volume decreases:
\begin{equation}\label{eq:dVdeta}
    \pdv{V_{neg}}{\eta} = \dfrac{1}{4\eta} \iint \limits_{W(x, y, \eta)<0} \nabla^2 W(x, y, \eta) dx dy
\end{equation}
Now, assuming that $\eta\to1$, we get the following formula for calculating vulnerability:
\begin{eqw}\label{eq:V_org}
    \mathbb{V} = \pdv{V_{neg}}{\eta}\Bigg|_{\eta=1}
    = \dfrac{1}{4} \iint \limits_{W(x, y)<0} \nabla^2 W(x, y) dx dy \,.
\end{eqw}

\section{Optimization of the vulnerability}\label{sec:optim_sq}

The Wigner function $W_{sq}$ of a squeezed state can be expressed in terms of the original Wigner function $W$ be the following replacement of coordinates:
\begin{eqw}
    W_{sq}(x, y) = W(x', y') ,
\end{eqw}
where
\begin{eqw}\label{eq:change_coordinates}
    \left(
    \begin{matrix}
        x' \\ y'
    \end{matrix}
    \right)
    =M(r, \phi)
    \left(
    \begin{matrix}
        x \\ y
    \end{matrix}
    \right)
\end{eqw}
and the matrix $M(r, \phi)$ can be represented as a composition of rotations and squeezing:
\begin{equation}\label{eq:M_def}
  M(r, \phi) = U^T S U
\end{equation}
with
\begin{eqw}
U &=
\left(
    \begin{matrix}
        \cos\phi && \sin\phi\\
        -\sin\phi && \cos\phi
    \end{matrix}
\right)\\
S &=
\left(
    \begin{matrix}
        e^{r} && 0\\
        0 && e^{-r}
    \end{matrix}
\right)
\end{eqw}
 Thus, vulnerability optimization reduces to the following minimization problem:
\begin{eqw}\label{eq:app_optim}
  \mathbb{V}_{\rm sqz} = \frac{1}{4}
    \iint\limits_{W_{sq} < 0} \left(\partial_x ^2 + \partial_y ^2\right) W_{sq}(x,y)\; dx\; dy = \min \,,
\end{eqw}
where $\mathbb{V}_{\rm sqz}$ is the value of $\mathbb{V}$ modified by the squeezing. Note that when replacing coordinates \eqref{eq:change_coordinates} in equation \eqref{eq:app_optim}, the Jacobian in is equal to one.

Introduce the factors $d_i$, which are directly related to the second derivatives of the original Wigner function in the regions of negativity:
\begin{eqw}
    d_i = \sigma_i^{jk} \int\limits_{W<0} \partial_j \partial_k W(x, y)\; dx\; dy \,,
\end{eqw}
where $\sigma_i$ are the Pauli matrices. Here and below, we use Einstein's notation for summation over repeated indexes, with $\partial_{1}$ and $\partial_{2}$ meaning $\partial_{x}$ and $\partial_{y}$ respectively.


Taking into account the following equation for the ``squeezed'' second derivatives:
\begin{eqw}
    \partial_{i} = M_i^j \partial'_j \,\Rightarrow \,
    \partial_x ^2 + \partial_y ^2 = \partial_{i}\partial^{i} = M_{i}^j M^i_k \partial'_j\partial'^k
\end{eqw}
the problem \eqref{eq:app_optim} boils down to optimizing the following expression (details in Appendix \ref{Appendix_B}):
\begin{eqw}\label{V_sqz_raw}
   \mathbb{V}_{\rm sqz}
   = \frac{1}{4}[d_0\cosh(2r)  + \left(d_3\cos (2\phi) + d_1\sin(2\phi)\right)\sinh(2r)] = \min \,.
\end{eqw}

Without loss of generality, we assume $r\geq 0$. In this case, straightforward minimization of $\mathbb{V}_{\rm sqz}$ in $r,\phi$ gives that the minimum is provided by
\begin{eqw}\label{eq:sqz_final}
    &e^{4r} = \frac{d_0 + \sqrt{d_1^2 + d_3^2}}{d_0 - \sqrt{d_1^2 + d_3^2}} \,,\\
    &\cos(2\phi) = -\frac{d_3}{\sqrt{d_1^2 + d_3^2}} \,, \\
    &\sin(2\phi) = -\frac{d_1}{\sqrt{d_1^2 + d_3^2}}
\end{eqw}
and is equal to
\begin{eqw}\label{eq:vulnerability_result}
  \mathbb{V}_{\rm sqz} = \frac{\sqrt{d_0^2 - d_1^2 - d_3^2}}{4}
\end{eqw}
For the comparison, the original no-squeezed vulnerability \eqref{eq:V_org} can be presented as follows:
\begin{equation}\label{V_org_d}
  \mathbb{V}_{\rm org} = \frac{d_0}{4} \,.
\end{equation}

Equations \eqref{eq:sqz_final} and \eqref{eq:vulnerability_result} are the main results of this paper and will be used below. We emphasize that squeezing makes it possible to improve the stability of the state, with the exception of the rare case where $d_1=d_3=0$. Since an optimally squeezed state cannot be made more stable by squeezing, the resulting state matrix $d_i\sigma^i$ is proportional to identity matrix.

\section{Examples}\label{sec:examples}

\subsection{General overview}

\begin{figure}
    \center
    \includegraphics[width=0.5\textwidth]{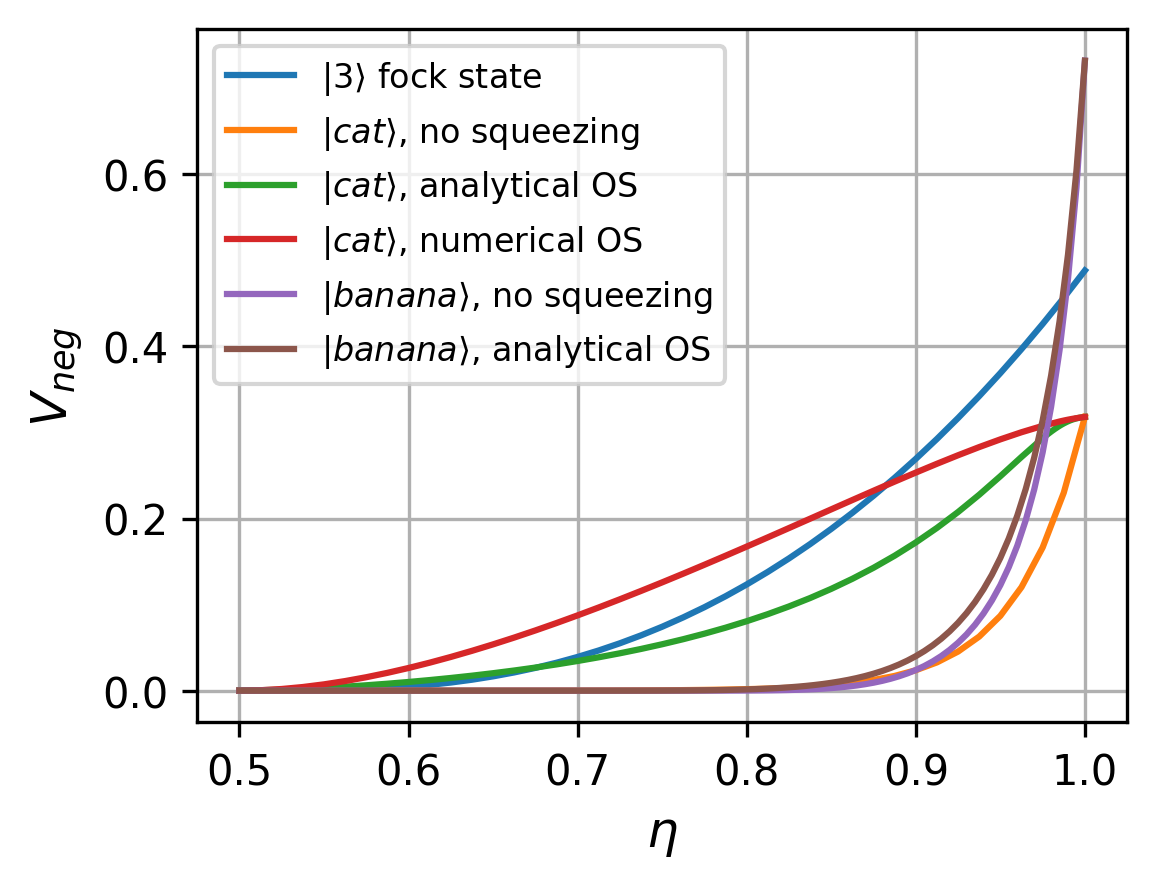}
    \caption{ The negativity volumes \eqref{eq:VnegDef} of typical quantum states on the quantum efficiency $\eta$. Blue line: Fock state $\ket{3}$. Orange line: Schrodinger cat state with $\alpha=3.6$. Green line: Schrodinger cat state with $\alpha=3.6$, pre-squeezed according to Eq.\,\ref{eq:sqz_final}. Red line: Schrodinger cat state with $\alpha=3.6$, pre-squeezed with the optimal squeeze parameters depending on $\eta$. Purple line: banana state at amplitude $\alpha=5$ and non-linearity $R=1.5$ (see notations in Sec.\,\ref{sec:banana}). Brown line: banana state with the same parameters, pre-squeezed according to Eq.\,\ref{eq:sqz_final}. }\label{fig:gr1}
\end{figure}

In Fig.\,\ref{fig:gr1}, dependences of the negativity volume  on losses are plotted for some typical quantum states. It can be seen from these plots that at $\eta\leq 1/2$ the negativity drops to zero \cite{leonhardt1994high}, since at $\eta=1/2$, the Wigner function after losses effectively turns to alway positive Husimi function. In the vicinity of this point, it grows polynomially \cite{nugmanov2024ustoichivost15242} and then increases monotonously with the increase of $\eta$.

{ It follows also from these plots that in the case of the non-squeezed quantum states, the fastest degradation of negativity, with the decrease of $\eta$, takes place in the area of small losses. This feature justifies the optimization procedure considered in Sec.\,\ref{sec:optim_sq}.}

At the same time, if the losses of $1-\eta$ in the system are not small, then the optimal squeezing parameters could differ from the ones given by Eqs.\,\eqref{eq:sqz_final}. To demonstrate this feature, we plotted also dependence of $V_{\rm neg}$ on $\eta$ for the squeezed Schrodinger cat case with the squeezing parameters was obtained as a result of local numerical optimization for each value $\eta$. {It follows from this plot that the dependence of the negativity on the quantum efficiency does not have to be convex.}

\subsection{Fock states}

\begin{figure}
    \center
    \includegraphics[width=0.5\textwidth]{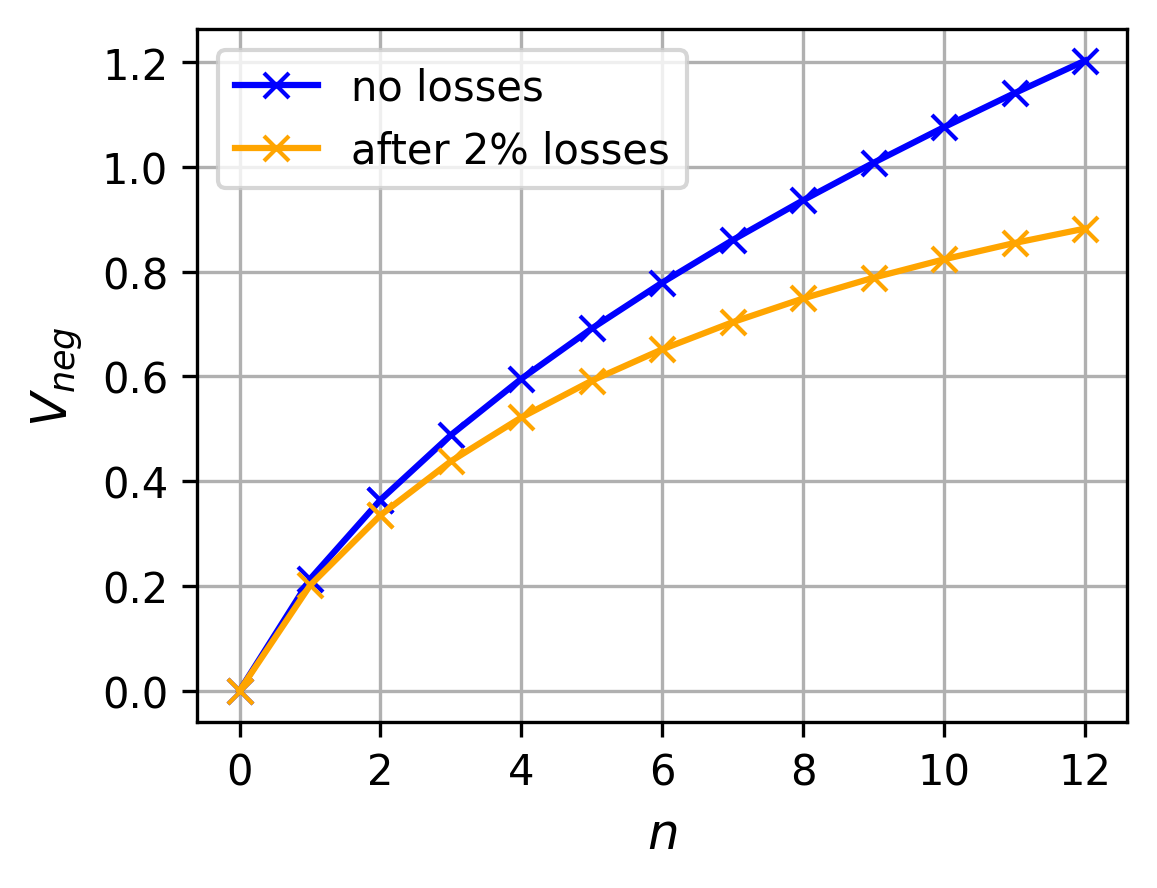}
    \caption{Negativity of the Fock state $\ket{n}$ as a function of $n$. As $n$ increases, so does the influence of losses.}\label{fig:gr2}
\end{figure}

\begin{figure}
    \center
    \includegraphics[width=0.5\textwidth]{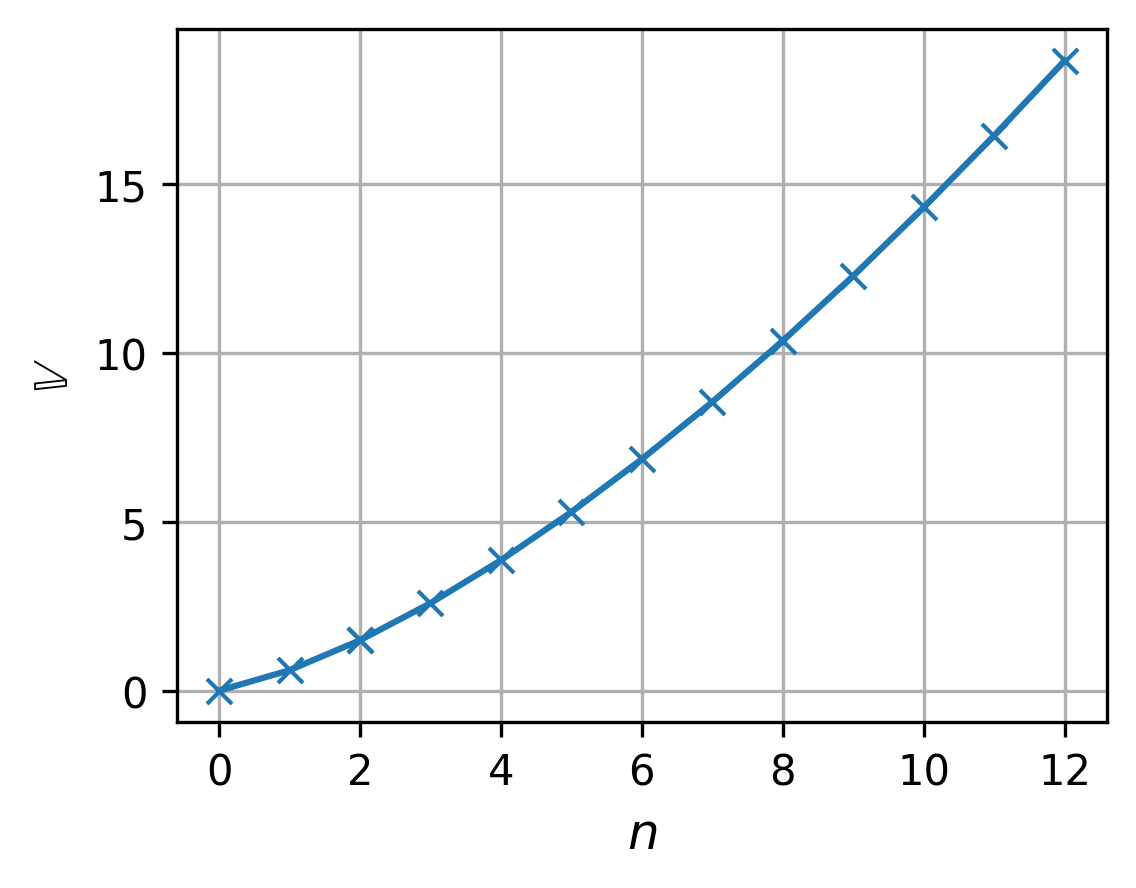}
    \caption{Nonlinear increase in the vulnerability of the Fock state $\ket{n}$ with the increase of $n$.}\label{fig:gr3}
\end{figure}

{Consider then} the Fock states. Fig.\,\ref{fig:gr2} shows that they have a very significant negativity.
This plot also shows that the impact of losses increases as $n$ photon number increases. { Fig.\,\ref{fig:gr3}, where dependence of $\mathbb{V}$ is plotted as a function of $n$, confirms this observation. Note that it is not possible to increase the robustness of these states using the to squeezing, because the Wigner functions of Fock states are radially symmetric and therefore any squeezing will only reduce their vulnerability to the losses.}

\subsection{Schrodinger's cat states}\label{sec:SC}

\begin{figure}
    \center
    \includegraphics[width=0.5\textwidth]{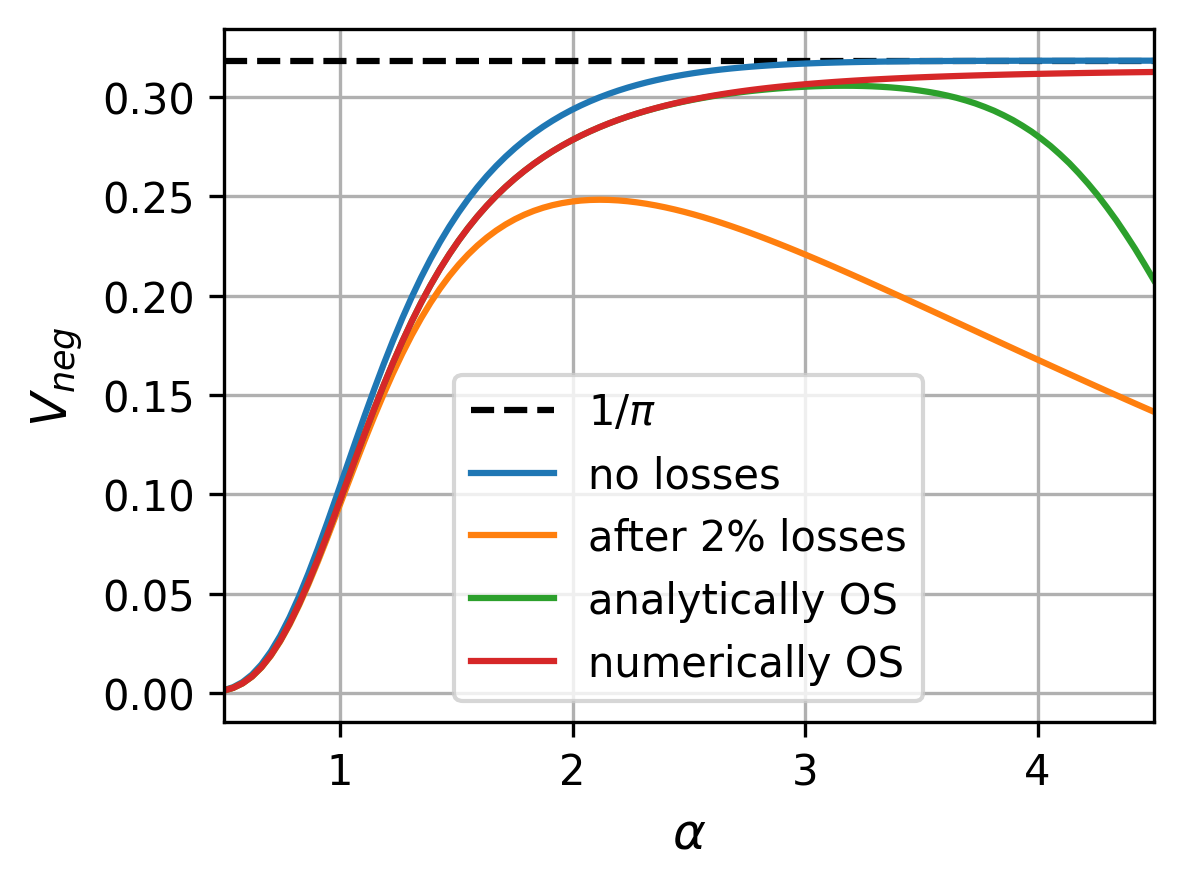}
    \caption{The negativity volumes \eqref{eq:VnegDef} of a Schrodinger's cat state as a functions of its amplitude $\alpha$. Dashed line: the theoretical limit on negativity at $\alpha\rightarrow\infty$. Orange line: the negativity with 2\% losses. Green line: with squeezing optimized using the formulae from the section \ref{sec:optim_sq}; note that it allows to keep more negativity up to $\alpha\sim 3$. Red line: the squeezing numerically optimized for 2\% losses. It is easy to see that it allows to preserve significantly more negativity for the same losses.
    }\label{fig:gr4}
\end{figure}
\begin{figure}
    \center
    \includegraphics[width=0.5\textwidth]{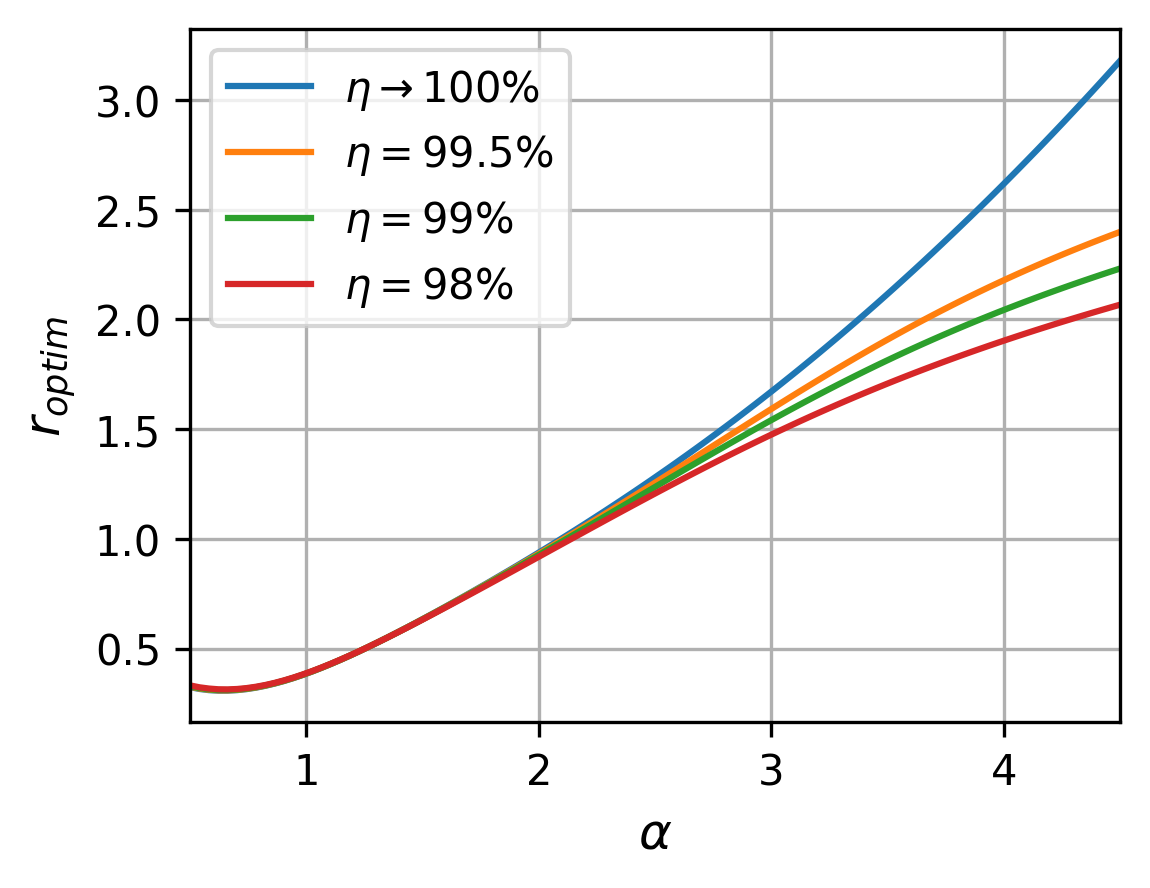}
    \caption{Dependence of the optimal squeezing parameter for the Schrödinger's cat state on the amplitude $\alpha$ for different loss values. Red line: the numerically optimized squeeze factor that was used in Fig.\,\ref{fig:gr4}. Blue line: the squeeze factor optimised using the formulae derived in Sec.\,\ref{sec:optim_sq}.}\label{fig:gr5}
\end{figure}
\begin{figure}
    \center
    \includegraphics[width=0.5\textwidth]{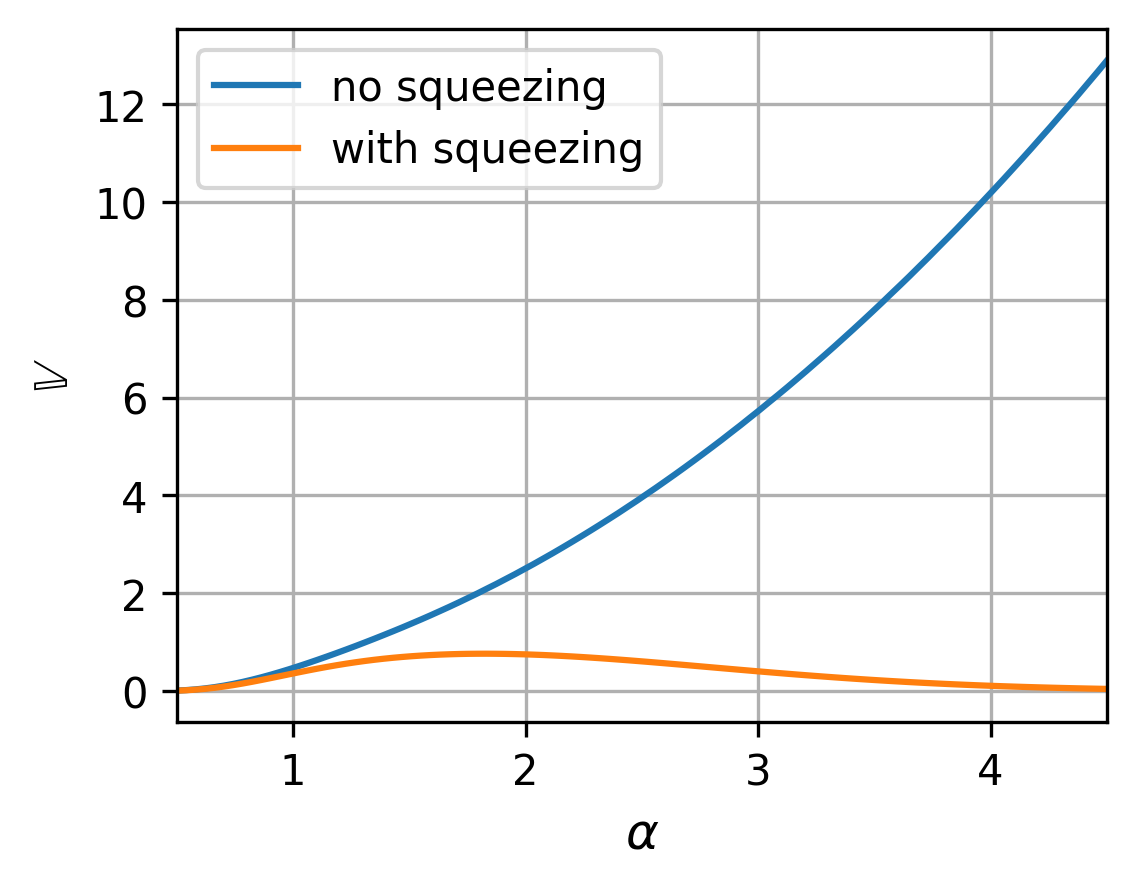}
    \caption{Vulnerability of the Schrodinger's cat state to losses $\mathbb{V}$ before and after applying optimal squeezing. Although it seems that the squeezing can help significantly reduce the impact of losses in the high alpha range, in this range, the higher corrections in the formula \eqref{eq:Vneg_by_eta} turn out to be significant, as we have seen from the graph \ref{fig:gr4}.}\label{fig:gr6}
\end{figure}

Now let's move on to considering the Schrodinger's cat state. Its negativity, unlike the Fock states, can be improved by squeezing, as it can be seen from Fig.\,\ref{fig:gr4}. The direction of optimal squeezing {is obvious from the symmetry of these states. It} coincides with the direction of the "interference fringes" in which negative values of the Wigner function are achieved.  It is important to note that Schrodinger's cat type states have a theoretical limit on maximum negativity. This is due to the fact that in the limit of $\alpha\rightarrow\infty$, in the vicinity of zero { ($x\sim y\to 0$)}, the Wigner function is a high-frequency sine modulated by a Gaussian function (see Appendix \ref{Appendix_C}). { We would like to note again that for given losses $\eta$, the numerically calculated} optimal squeezing may differ from the optimal squeezing for low losses. {This effect is clearly visible in Figs.\,\ref{fig:gr4} and \ref{fig:gr5}.
It can be explained as follows.} Although the first derivative of the negativity volume can be greatly reduced by squeezing, higher corrections in formula \eqref{eq:Vneg_by_eta} can make a significant contribution.

\begin{figure}
    \center
    \includegraphics[width=0.5\textwidth]{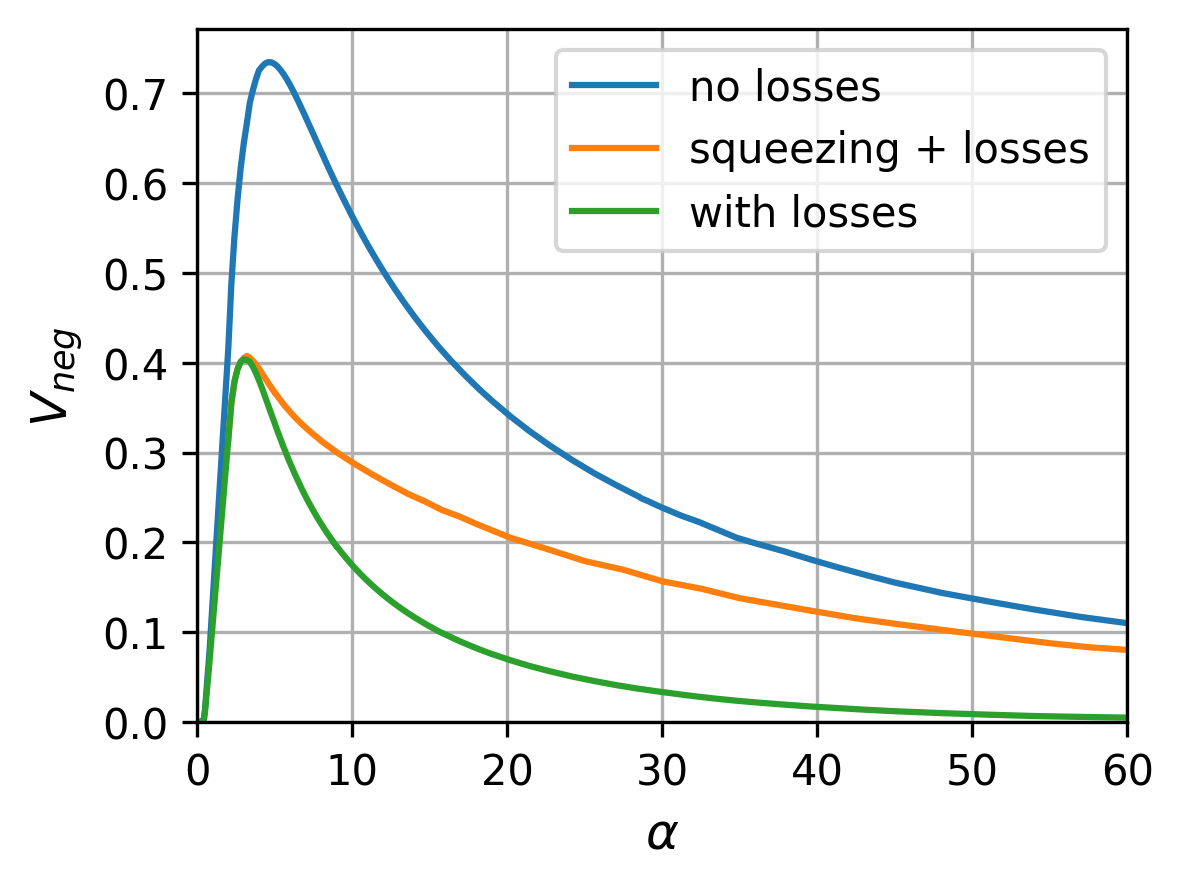}
    \caption{Volume of negativity of the banana state as a function of the parameter $\alpha$ for the fixed non-linearity parameter $R=1.5$. }\label{fig:gr7}
\end{figure}
\begin{figure}
    \center
    \includegraphics[width=0.5\textwidth]{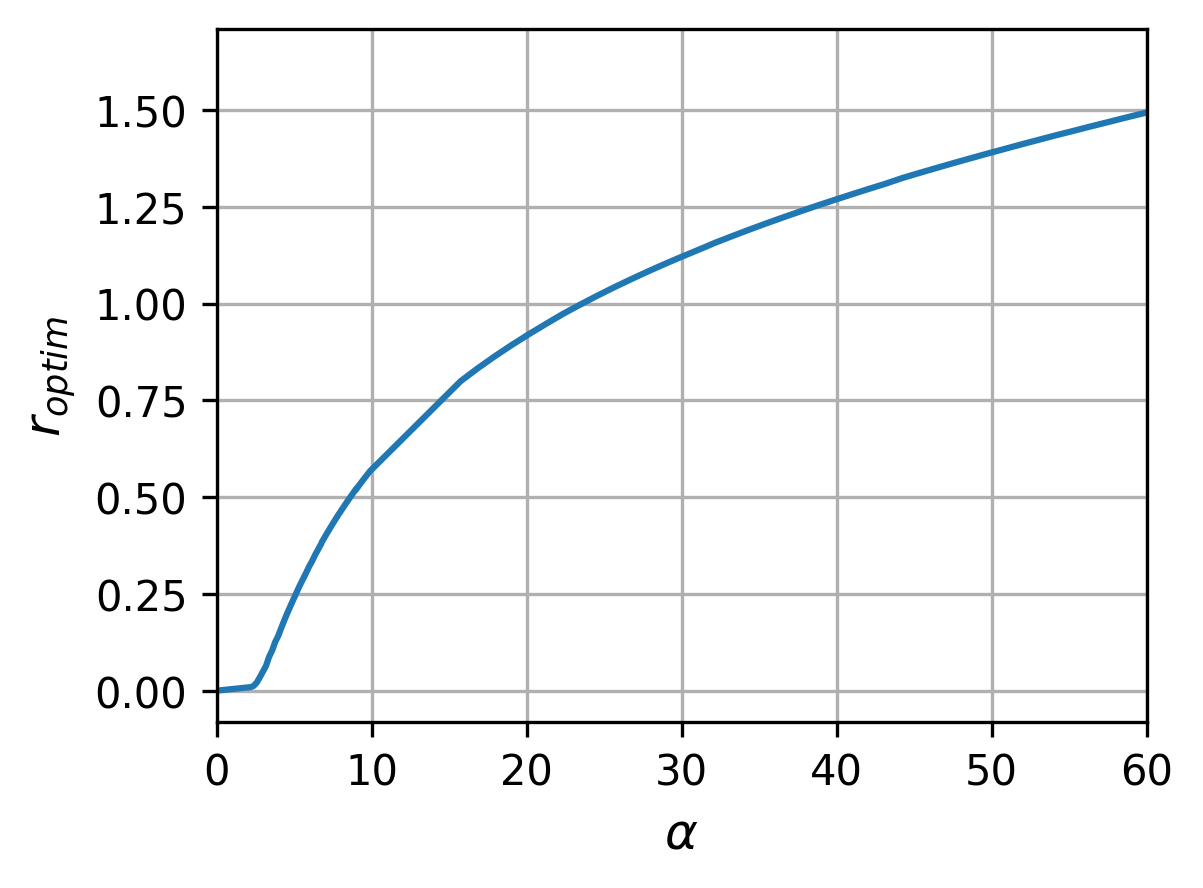}
    \caption{ Optimal squeeze factor defined by Eq.\,\eqref{eq:sqz_final} as a function of the parameter $\alpha$ for the fixed non-linearity parameter $R=1.5$.}\label{fig:gr8}
\end{figure}
\begin{figure}
    \center
    \includegraphics[width=0.5\textwidth]{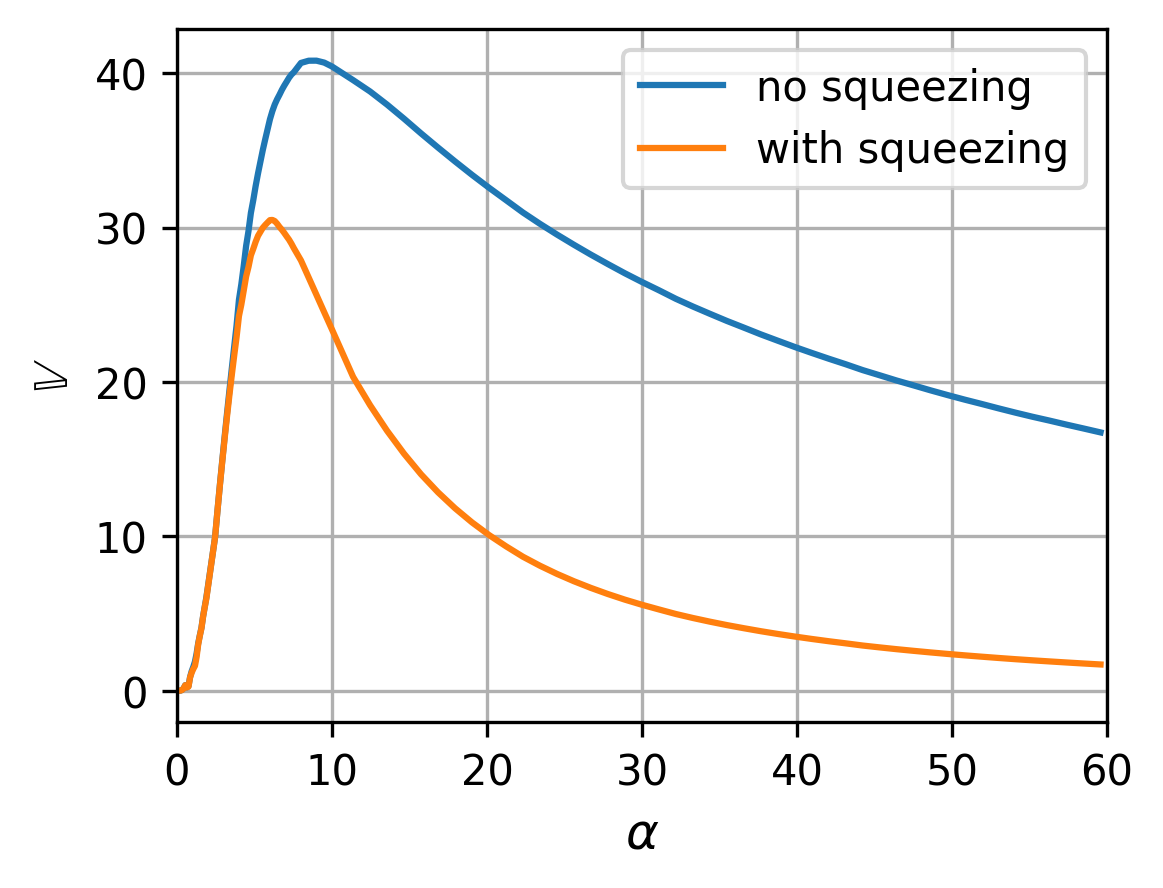}
    \caption{ Vulnerability $\mathbb{V}$ of the banana states as a function of $\alpha$ for the fixed non-linearity parameter $R=1.5$. It can be seen that $\mathbb{V}$ decreases fast with the increase of $\alpha$.}\label{fig:gr9}
\end{figure}

\subsection{Banana states}\label{sec:banana}

Finallyy, consider the so-called banana state:
\begin{eqw}
    \ket{banana} = e^{-i \Gamma\hat{n}^2}\ket{\alpha} \,,
\end{eqw}
{where $\Gamma$ is the Kerr nonlinearity factor. The name ``banana state''} was introduced by Kitagawa in his work \cite{kitagawa1986number}. It originates from the fact that the level lines of quasi-probability distributions of this state remind a banana.

Optical cubic nonlinearity is relatively small, therefore, to observe significant nonlinear effects and negativity in particular, it is necessary to use initially highly excited Gaussian states, that is, a large values of the parameter $\alpha$.

As it was noted in Ref.\,\cite{nougmanov2024effective}, {where the effective method of calculation of the Wigner function of banana state used here was developed,} the main parameter indicating the degree of manifestation of nonlinearity is $R=\alpha^2\Gamma$, therefore in this paper we will compare states with the same value of $R=1.5$. The negativity of this state is significant enough, as can be seen in Fig.\,\ref{fig:gr7}. In Figs.\,\ref{fig:gr8} and \ref{fig:gr9} it can be seen also that using the squeezing, the vulnerability $\mathbb{V}$ of the ``banana'' states can be significantly reduced. However, numerical calculations show that neither optimal "analytical" squeezing nor numerically {calculated locally optimal squeezing allow} to preserve at least some noticeable part of the negativity. The addition of even very small losses, like of tenths of a percent, destroys almost all negativity. This result means that in optics, observation of the quantum properties of the banana state is fraught with significant difficulties.

\section{Conclusion}\label{sec:conclusion}
Using the volume of negativity of the Wigner function as a measure of the quantumness we showed that pre-squeezing can reduce its vulnerability to losses. We found an analytical expression for optimal squeezing parameters \eqref{eq:sqz_final}. Considering Schrodinger's cat state and banana states, we showed that optimal pre-squeezing can significantly increase the resistance of states to losses. The vulnerability of Fock states cannot be reduced, due to axial symmetry of the states. Plots of the negativity of various states (Fig. \ref{fig:gr1}) demonstrate that the largest part of the volume of negativity is lost at small losses, which explains determining the vulnerability of states \eqref{eq:V_org}.
\section*{Acknowledgments}
This work was supported by the Russian Science Foundation (Project No. 25-12-00263).
\section{Appendix}
\subsection{First derivative of $V_{neg}(\eta)$}\label{Appendix_A}
Let's relate the derivative of the Wigner function by $\eta$ to its local behavior.
In order to link derivatives by losses and coordinates, it is necessary to study the behavior of the kernel of the integral transformation \eqref{eq:W_eta}. For this purpose, it is convenient to temporarily introduce the notation:
\begin{eqw}
    f(\eta, x, y) = \br{x-\sqrt{\eta}\tilde x}^2+\br{y-\sqrt{\eta}\tilde y}^2
\end{eqw}
Now we can take the derivative of kernel as if $f$ were an arbitrary function and substitute the entered notation with the next line:
\begin{eqw}
    \pdv{\eta}\frac{\exp\br{-\frac{f(\eta, x, y)}{1-\eta}}}{(1-\eta)} &= -\frac{\exp\br{-\frac{f(\eta, x, y)}{1-\eta}}}{(1-\eta)^3}\br{f(\eta, x, y) + (1-\eta)\br{f'_{\eta}(\eta, x, y) -1}}\\
    &= \frac{\exp\br{-\frac{f(\eta, x, y)}{1-\eta}}}{(1-\eta)^3}\br{\eta-1 + x^2 + \tilde{x}^2 + y^2 + \tilde{y}^2 - \frac{\br{x\tilde{x}+y\tilde{y}}(1+\eta)}{\sqrt{\eta}}}
\end{eqw}
We will do the same with the following construction:
\begin{eqw}
    \br{ x \pdv{x}  + \frac{1}{2}\frac{\partial^2}{\partial x^2}}\frac{\exp\br{-\frac{f(\eta, x, y)}{1-\eta}}}{(1-\eta)}&=\frac{\exp\br{-\frac{f(\eta, x, y)}{1-\eta}}}{(1-\eta)^2}\br{x f'_x(\eta, x, y)+\frac{f''_{xx}(\eta, x, y)}{2} - \frac{f'^2_x(\eta, x, y)}{2(1-\eta)}}\\
    &= \frac{\exp\br{-\frac{f(\eta, x, y)}{1-\eta}}}{(1-\eta)^3}\br{1-\eta+2 x \tilde{x}(1+\eta) -2\eta\br{x^2+\tilde{x}^2} }
\end{eqw}
Combining linear combinations of the obtained derivatives and using symmetry with respect to $x$ and $y$, we get following expression \cite{nugmanov2024ustoichivost15242}:
\begin{equation}\label{eq:Wf_decay_rate}
    \pdv{\eta} W(x, y,\eta) = -\dfrac{1}{\eta} \biggl[1 +\frac{1}{2}\left(x\frac{\partial}{\partial x}+y \frac{\partial}{\partial y}\right)
    + \frac{1}{4}\left(\frac{\partial^2}{\partial x^2}+\frac{\partial^2}{\partial y^2}\right) \biggr] W(x, y,\eta).
\end{equation}
Substituting this expression into definition of negativity volume \eqref{eq:VnegDef} and using integration by parts, we obtain an explicit expression for the derivative of negativity at any $\eta$:
\begin{equation}\label{eq:dVdeta}
    \pdv{V_{neg}}{\eta} = \dfrac{1}{4\eta} \iint \limits_{W(x, y, \eta)<0} \nabla^2 W(x, y, \eta) dx dy
\end{equation}
\subsection{Vulnerability of squeezed state}\label{Appendix_B}
In this appendix, we introduce a matrix of integrals from the original Wigner function:
\begin{eqw}
    L_{ij} =
      \iint\limits_{W < 0} \partial_i'\partial_{j}' W(x',y')\; dx'\; dy
\end{eqw}
Let's replace the coordinates in equation \eqref{eq:app_optim}. When switching from squeezed coordinates to unsqueezed coordinates, $W_{sq}(x, y) = W(x', y')$ is performed.
\begin{eqw}
    \mathbb{V}_{\rm sqz} = \frac{1}{4}
      \iint\limits_{W < 0} M_{i}^j M^i_k \partial'_j\partial'^k W(x',y')\; dx'\; dy' = \frac{1}{4} M_{i}^j M^i_k L_{j}^k = \frac{1}{4} \Tr\br{M^2 L}
\end{eqw}
Now we use the expression \eqref{eq:M_def} for the matrix $M$ and decompose it into Pauli matrices:
\begin{eqw}
    M^2 = U^{T}S^2 U = \sigma_0 \cosh(2r) + \br{\sigma_3 \cos(2\phi) -\sigma_1\sin(2\phi)}\sinh(2r)
\end{eqw}
Thus, we come to equation \eqref{V_sqz_raw}.

\subsection{Schrödinger cat maximum negativity}\label{Appendix_C}
The Wigner-Schrodinger function is expressed as follows \cite{schleich2015quantum}:
\begin{eqw}
    W(x, y) &= \frac{1}{1+e^{-2\alpha^2}}\br{W_{-}(x, y) + W_{+}(x, y) + W_{int}(x, y)}\\
    W_{\pm} &= \frac{1}{2\pi}\exp\br{-y^2-\br{x\pm\sqrt{2}\alpha}^2}\\
    W_{int} &= \frac{1}{\pi}\exp\br{-y^2-x^2}\cos\br{2\sqrt{2}\alpha y}
\end{eqw}
When calculating the volume of negativity at $\alpha\rightarrow\infty$, the main contribution is made by $W_{int}$, while the influence of $W_{\pm}$ can be ignored in areas where negativity is concentrated:
\begin{eqw}
    V_{neg} = \frac{1}{2} \iint \frac{\abs{W(x, y)} - W(x, y)}{2} dx \;dy \approx  \frac{1}{2} \iint \frac{\abs{W_{int}(x, y)} - W_{int}(x, y)}{2} dx \;dy
\end{eqw}

Using the Fourier series expansion, we can find the following expression:
\begin{eqw}
    \frac{\abs{\cos(x)} - \cos(x)}{2} = \frac{1}{\pi} -\frac{\cos(x)}{2}+ \frac{2}{\pi}\sum_{n=1}^{\infty} (-1)^n\frac{\cos(2 n x)}{1-4 n^2}
\end{eqw}
Keeping only the zero contribution in the expansion, we get that $V_{neg}\rightarrow1/\pi$

\end{document}

%% file: preambule.tex
\usepackage[T2A]{fontenc} 
\usepackage[utf8]{inputenc} 
\usepackage[english]{babel}
\usepackage[left=2cm,right=2cm,top=2cm,bottom=3cm,bindingoffset=0cm]{geometry}
\usepackage{indentfirst}
\usepackage{graphicx}
\graphicspath{}
\DeclareGraphicsExtensions{.pdf,.png,.jpg,.jpeg}

\usepackage[table,xcdraw]{xcolor}
\usepackage{booktabs}

\usepackage{hyperref}
\usepackage{physics}
\usepackage{amssymb, amsmath}

\usepackage{pgfplots}
\pgfplotsset{compat=1.15}
\usepackage{mathrsfs}
\usetikzlibrary{arrows}

\usepackage{mathtools}
\DeclarePairedDelimiter\autobracket{(}{)}
\newcommand{\br}[1]{\autobracket*{#1}}

\usepackage{empheq}
\newenvironment{eqw}{\begin{equation} \begin{aligned}}{\end{aligned}    \end{equation}}
\newenvironment{eqw*}{\begin{equation*} \begin{aligned}}{\end{aligned}    \end{equation*}}

%% file: main_FK2.bbl
\begin{thebibliography}{10}

\bibitem{schleich2015quantum}
Wolfgang~P Schleich.
\newblock {\em Quantum optics in phase space}.
\newblock John Wiley \& Sons, 2015.

\bibitem{bohr1928quantum}
Niels Bohr et~al.
\newblock {\em The quantum postulate and the recent development of atomic theory}, volume~3.
\newblock Printed in Great Britain by R. \& R. Clarke, Limited, 1928.

\bibitem{hudson1974wigner}
Robin~L Hudson.
\newblock When is the wigner quasi-probability density non-negative?
\newblock {\em Reports on Mathematical Physics}, 6(2):249--252, 1974.

\bibitem{bell2004speakable}
John~Stewart Bell.
\newblock {\em Speakable and unspeakable in quantum mechanics: Collected papers on quantum philosophy}.
\newblock Cambridge university press, 2004.

\bibitem{mari2012positive}
Andrea Mari and Jens Eisert.
\newblock Positive wigner functions render classical simulation of quantum computation efficient.
\newblock {\em Physical review letters}, 109(23):230503, 2012.

\bibitem{walschaers2021non}
Mattia Walschaers.
\newblock Non-gaussian quantum states and where to find them.
\newblock {\em PRX quantum}, 2(3):030204, 2021.

\bibitem{Gorshenin_LPL_21_065201_2024}
VL~Gorshenin.
\newblock Using schr{\"o}dinger cat quantum state for detection of a given phase shift.
\newblock {\em Laser Physics Letters}, 21(6):065201, 2024.

\bibitem{Gorshenin_JOSAB_42_425_2025}
VL~Gorshenin.
\newblock Preparation of the schr{\"o}dinger cat quantum state using parametric down-conversion interaction.
\newblock {\em Journal of the Optical Society of America B}, 42(2):425--430, 2025.

\bibitem{Zurek_RMP_75_715_2003}
Wojciech~Hubert Zurek.
\newblock Decoherence, einselection, and the quantum origins of the classical.
\newblock {\em Rev. Mod. Phys.}, 75:715--775, May 2003.

\bibitem{grangier1987squeezed}
Philippe Grangier, RE~Slusher, B~Yurke, and A~LaPorta.
\newblock Squeezed-light--enhanced polarization interferometer.
\newblock {\em Physical review letters}, 59(19):2153, 1987.

\bibitem{xiao1987precision}
Min Xiao, Ling-An Wu, and H~Jeffrey Kimble.
\newblock Precision measurement beyond the shot-noise limit.
\newblock {\em Physical review letters}, 59(3):278, 1987.

\bibitem{ligo2011gravitational}
A gravitational wave observatory operating beyond the quantum shot-noise limit.
\newblock {\em Nature Physics}, 7(12):962--965, 2011.

\bibitem{aasi2015advanced}
Junaid Aasi, BP~Abbott, Richard Abbott, Thomas Abbott, MR~Abernathy, Kendall Ackley, Carl Adams, Thomas Adams, Paolo Addesso, RX~Adhikari, et~al.
\newblock Advanced ligo.
\newblock {\em Classical and quantum gravity}, 32(7):074001, 2015.

\bibitem{ralph1999continuous}
Timothy~C Ralph.
\newblock Continuous variable quantum cryptography.
\newblock {\em Physical Review A}, 61(1):010303, 1999.

\bibitem{martin2016bell}
J{\'e}r{\^o}me Martin and Vincent Vennin.
\newblock Bell inequalities for continuous-variable systems in generic squeezed states.
\newblock {\em Physical Review A}, 93(6):062117, 2016.

\bibitem{yuen1978optical}
Horace Yuen and J~Shapiro.
\newblock Optical communication with two-photon coherent states--part i: Quantum-state propagation and quantum-noise.
\newblock {\em IEEE Transactions on Information Theory}, 24(6):657--668, 1978.

\bibitem{leonhardt1994high}
U~Leonhardt and H~Paul.
\newblock High-accuracy optical homodyne detection with low-efficiency detectors:" preamplification" from antisqueezing.
\newblock {\em Physical review letters}, 72(26):4086, 1994.

\bibitem{kenfack2004negativity}
Anatole Kenfack and Karol {\.Z}yczkowski.
\newblock Negativity of the wigner function as an indicator of non-classicality.
\newblock {\em Journal of Optics B: Quantum and Semiclassical Optics}, 6(10):396, 2004.

\bibitem{nugmanov2024ustoichivost15242}
B.~N. Nugmanov.
\newblock Resistance of the negativity of the wigner function to dissipation.
\newblock {\em Memoirs of the Faculty of Physics}, (3), 2024.
\newblock \url{http://publish.phys.msu.ru/search/get?f62563}.

\bibitem{kitagawa1986number}
M~Kitagawa and Y~Yamamoto.
\newblock Number-phase minimum-uncertainty state with reduced number uncertainty in a kerr nonlinear interferometer.
\newblock {\em Physical Review A}, 34(5):3974, 1986.

\bibitem{nougmanov2024effective}
Boulat Nougmanov.
\newblock Effective algorithms for calculation of quasi-probability distributions of bright “banana” states.
\newblock {\em Journal of the Optical Society of America B}, 41(7):1573--1586, 2024.

\end{thebibliography}
